\documentclass{iaus}
\usepackage{graphicx}

\title[The resonant binary BI\,108] 
{The resonant B1II\,+\,B1II binary BI\,108\footnote{Based on observations
    under ESO program 382.D-0311.}}

\author[Th.\ Rivinius, R.~E.~Mennickent, Z.~Ko{\l}aczkowski]   
{Th.\ Rivinius$^{1}$, R.~E.~Mennickent$^2$, Z.~Ko{\l}aczkowski$^3$ 
}

\affiliation{$^1$ESO, Chile; 
$^2$U.~de Concepc\'ion, Chile;
$^3$U.~Wroclawskiego, Poland
}

\pubyear{2011}
\volume{272}  
\pagerange{1--2}
\jname{Active OB stars: structure, evolution, mass loss and critical limits}
\editors{C.\ Neiner, G.\ Wade, G.\ Meynet \& G.\ Peters, eds.}
\begin{document}
\maketitle

\begin{abstract}
BI\,108 is a luminous variable star in the Large Magellanic Cloud classified
B1\,II. The variability consists of two resonant periods (3:2), of which only
one is orbital, however. We discuss possible mechanisms responsible for the
second period and its resonant locking.

\keywords{stars: binaries, stars: early-type}
\end{abstract}

\firstsection 
\section{Introduction}
The observed period of BI\,108 (OGLE-ID: LMC SC9-125719, MACHO-ID 79.5378.25)
of 10.73\,d has {six equidistant but distinct minima} and some symmetry
  around the deepest minimum.  At close inspection, it became clear that the
  lightcurve can actually be disentangled into two periods with a resonant
  ratio of 3:2 (see figures in \cite[Ko{\l}aczkowski et al.\ (in
    press)]{brnoconf}). From October 2008 to January 2009 20 spectra were
  taken at {ten epochs} with the echelle {spectrograph UVES}, mounted at the
  8.2\,m telescope UT2 on Cerro Paranal.

\section{Orbital parameters and spectral disentangling}
The obtained spectra are of an SB2 composite nature, however, only one single
period is present in the radial velocities, namely $P_{\rm
  orb.}=5.37$\,d. Star A and B have an almost circular orbit, similar mass,
and are in a similar evolutionary stage, being both very early B supergiants,
B0\,II+B0\,II. No circumstellar gas, i.e.\ no mass transfer, was detected.
The spectral disentangling was done with VO-KOREL based on KOREL,
\cite[Hadrava (1995)]{had95}: \medskip

\noindent{\footnotesize
\begin{center}
\begin{tabular}{lr|lr}
$P_{\rm sup.}$ & 10.73309\,d  \hspace*{5mm}&\hspace*{5mm} Periastron long. &
  93$^\circ$ \\

$P_{\rm res.}$ & 3.57793\,d  \hspace*{5mm}&\hspace*{5mm}              $K_1$ & 170\,km/s \\            
$P_{\rm orb.}$ & 5.36654\,d  \hspace*{5mm}&\hspace*{5mm}              $K_2$ & 225\,km/s \\            
Epoch (min. light) & MJD=51\,163.3915  \hspace*{5mm}&\hspace*{5mm}    $q$   & 0.76 \\                 
Periastron date & MJD=54\,742.8345  \hspace*{5mm}& \hspace*{5mm}      $a \sin i$ & 41.5\,R$_\odot$\\  
Eccentricity & 0.08  \hspace*{5mm}&\hspace*{5mm}                      $M \sin^3 i$ & 33.6\,M$_\odot$\\
\end{tabular}
\end{center}}
\vspace*{5mm}

The {resonant period} is not completely absent in the spectra, however, but
{modulates the line strengths}. In particular, the total equivalent width (EW)
over both components is about constant, while each component varies strongly,
i.e.\ it seems as if a certain fraction of {equivalent width is exchanged}
between the stars with a period of $P_{\rm res.}=3.58$\,d, although admittedly
our data is too scarce to claim so with much certainty. For brevity, we call
this behavior ``EW-shuffling'' (see Fig.~1, right) in the following.

\begin{figure}
\centering
\parbox{\textwidth}{%
\parbox{0.5\textwidth}{\includegraphics[angle=270,width=0.5\textwidth,clip]{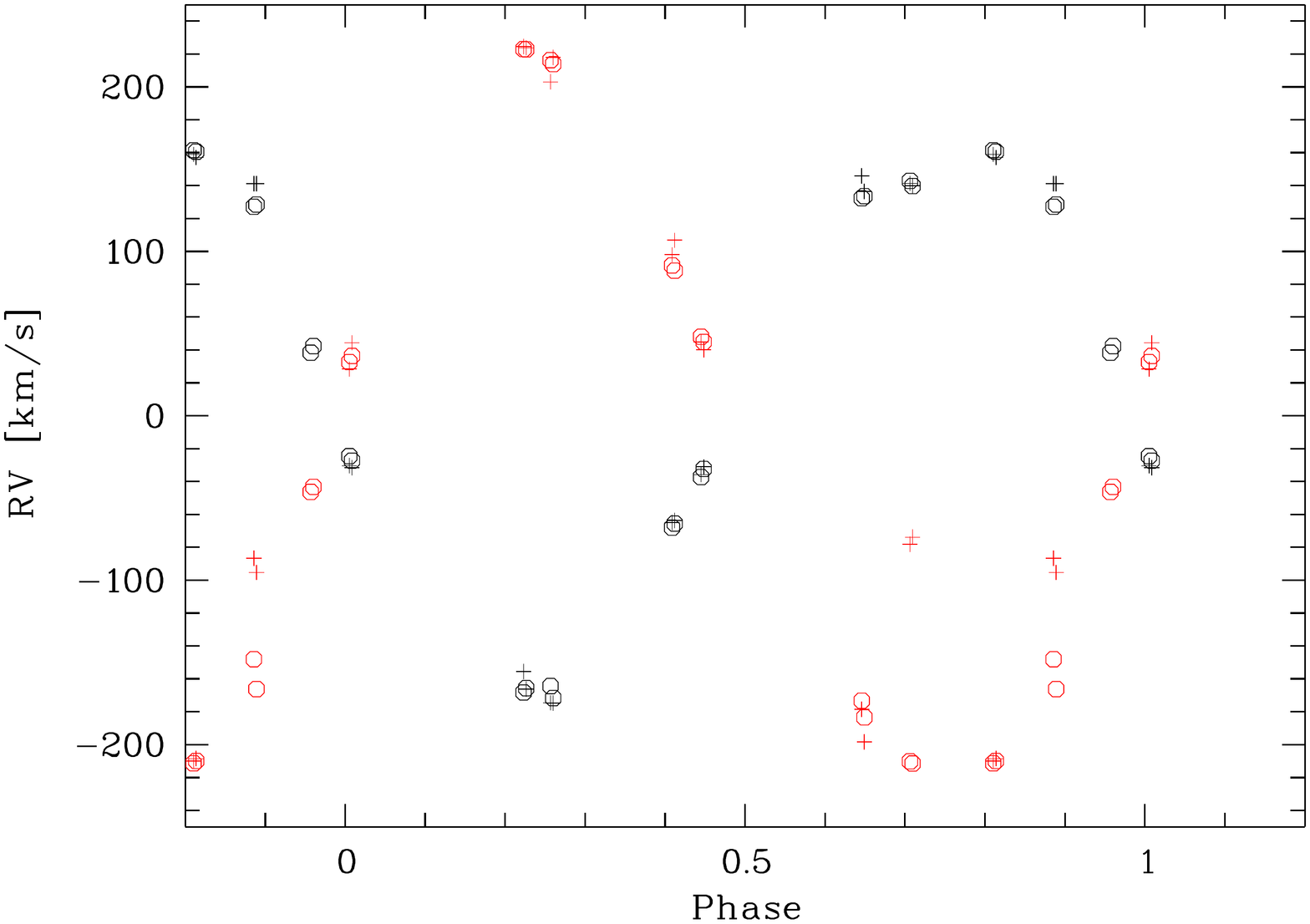}}%
\parbox{0.5\textwidth}{\includegraphics[angle=270,width=0.5\textwidth,clip]{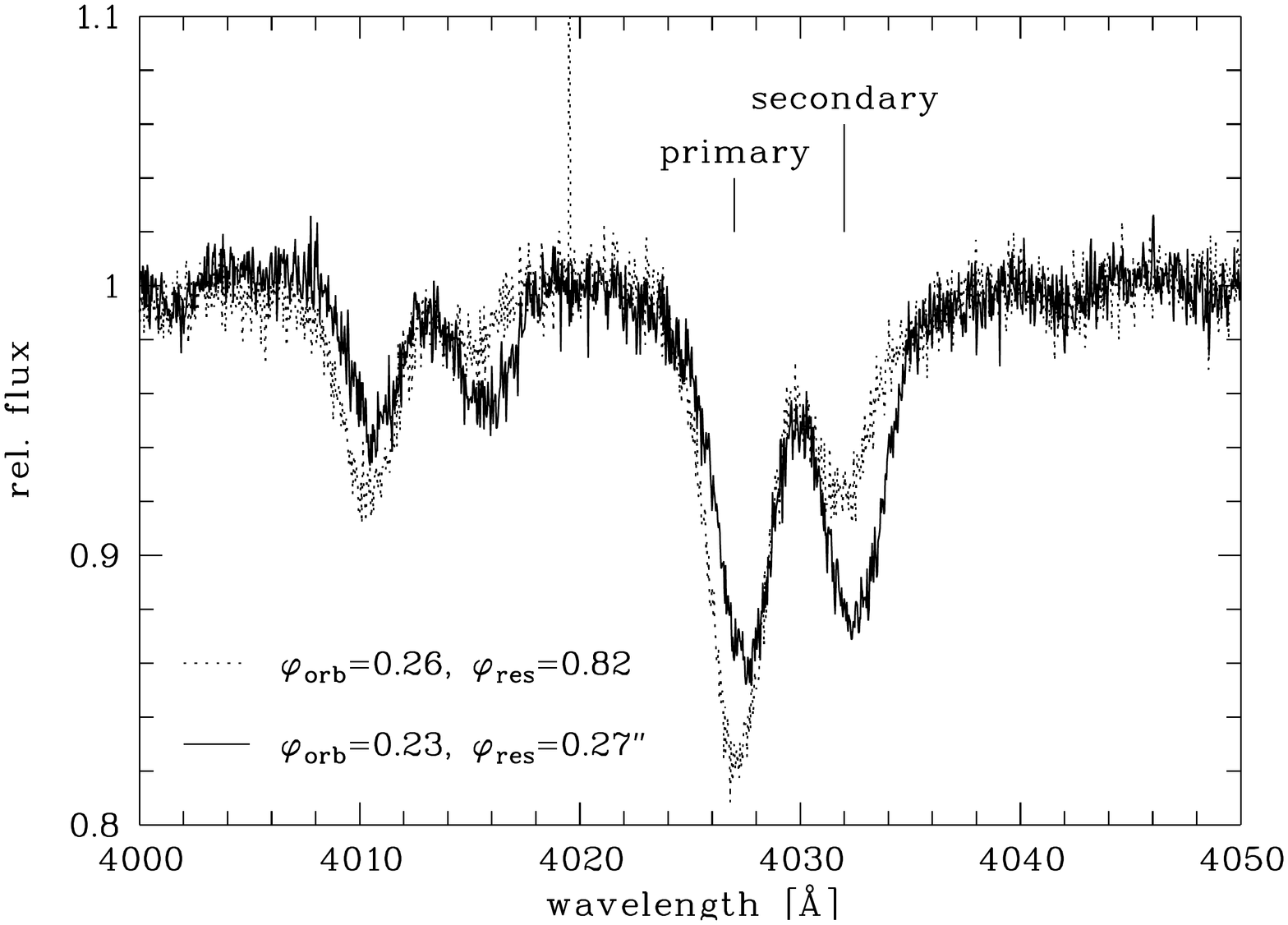}}%
}
\caption[]{
\centering
Left panel: Orbital velocities determined by VO-KOREL
   disentangling. Crosses are values derived from Si{\sc iii}\,4553, circles
   from H$\beta$. Sort with $P_{\rm orb.}$=5.36654\,d.
\newline Right panel: UVES spectra taken at similar orbital, but opposite
resonant quadrature phase. The ``EW shuffling'' between both lines is
obvious.}
\label{FNCMa}
\end{figure}

\section{Discussion}

A good hypothesis for the explanation of the system has to fulfill a number of
criteria:\\[0.5mm]
{\bf 1.)} Explain why the superperiod is disentangled into two {resonant
  periods}.  {\bf 2.)} Provide a strong and stable locking mechanism between
the two periods.  {\bf 3.)} Explain the phase relation, i.e. {\bf a.)}
$T_{\rm min, res.} = T_{\rm min, orb.}$, and {\bf b.)} the RV vs.\ light curve
relations, not as a line-of-sight coincidence, but for all aspects.  {\bf 4.)}
Provide an {evolution scenario}, in which the current system, hosting (at
least) two B0\,II stars, could have evolved {from a plausible off-the-shelf
  system}.  {\bf 5.)} Provide at least a toy model for the EW shuffling.

In the following we discuss several hypothesis:
{\bf Lagrangian triple:} A
    Lagrangian triple is a system with three stars A+B+C on a common orbit,
    trailing each other with $\Delta\phi = 120^\circ$. In a more general
    scheme, concentric orbits for stars with unequal masses are possible as
    well. EW shuffling might be due to the unrecognized component switching
    between A and B. However, such a system quite {strongly violates
      criteria 2, 3b, and 4.} 
{\bf Hierarchical system:} A system of type (Aa+Ab)+(Ba+Bb) was the original
suggestion when the system was first published \cite[Ofir, 2008]{ofi08}.
However, this suggestion falls short in criteria 1, 2, 3a, and 3b.
{\bf Two-star magnetic resonance:} This hypothesis assumes only two stars A+B,
with the spin-orbit relation locked by magnetic fields. Criteria 1 and 3 are
{naturally fulfilled} and 4 is not a major problem a priori, but the {3:2}
locking ratio might be an issue for criterion 2 (at least for {dipoles}). Yet
neither the {EW shuffling} nor the photometric curve look like {anything a
  magnetic star} usually has to offer.
{\bf Two-star tidal resonance:} Also only two stars A+B, but their spin-orbit
relation are locked due to tidal interaction. Again criteria 1 and 3 are
{naturally fulfilled}, while according to \cite[Witte \& Savonije
  (2001)]{wit01} 2 and 4 are {plausible}. Geometrical
  distortion and light modulation by a strong tidal wave is {at least a toy
  model} for the EW behaviour.

As Witte \& Savonije (2001) point out in a study of a
10\,M$_\odot$+10\,M$_\odot$ main sequence binary, such a tidal resonant
locking might actually be quite common and stable during extended phases of
the orbital circularization. Since we deal with massive stars, it is as well
plausible that the stars have {evolved} away from the main sequence { before
  being fully circularized}.  Although the tidally locked scenario leaves
uncomfortably many open questions, it is the one requiring the least extreme
assumptions, satisfying Occam's razor.


\begin{thebibliography}{}

\bibitem[Ko{\l}aczkowski et al., (in press)]{brnoconf} {Ko{\l}aczkowski,
  Mennickent, Rivinius} in ``Binaries: Key to Comprehension of the Universe''
  held 2009 in Brno, in press, arXiv:1004.5464.

\bibitem[Hardava, (1995)]{had95}{Hadrava} 1995, A\&AS 114, 393

\bibitem[Ofir, (2008)]{ofi08}{Ofir} 2008, IBVS 5868

\bibitem[Witte \& Savonije, (2001)]{wit01}{Witte \& Savonije} 2001, A\&A 366,
  840

\end{thebibliography}
\end{document}